# A novel acquisition method of nuclear spectrum based on pulse area analysis[*]

Li Dongcang(李东仓)[1,†], Ren Zhongguo(任忠国)[1, 2], Yang Lei(杨磊)[1], Qi Zhong(祁中)[1],

Meng Xiangting(孟祥厅)[1], Hu Bitao(胡碧涛)[1]

（1. School of Nuclear Science and Technology, Lanzhou University, Lanzhou 730000, China

2．China Academy of Engineering Physics, Mianyang 621907, China）

**Abstract:** A novel simple method based on pulse area analysis(PAA) was presented for acquisition nuclear spectrum by the digitizer. PAA method can be used as a substitute for the traditional method of pulse height analysis (PHA). In the PAA method a commercial digitizer was employed to sample and sum in the pulse, and the area of pulse is proportional to the energy of the detected radiation. The results of simulation and experiment indicate the great advantages of PAA method, especially as the count rate is high and shaping time constant is small. When shaping time constant is 0.5μs, the energy resolution of PAA is about 66% better than that of PHA.

## 1 Introduction

PHA is a traditional method of nuclear spectrum acquisition, in which a pulse from a detector is transformed, amplified, shaped and filtered by the preamplifier and the pulse amplifier. The suitable pulses enter the acquisition system and the spectrum of pulse height distribution can be gotten by this way. The general ADC (analog to digital converter) can't directly be used in PHA because of its poor differential nonlinearity (DNL). To decrease the disadvantage the DNL of ADC should be reduced by some special way, such as the slide scales[1].

Recently, a great progress has been made in digital processing technique. Faster DSPs, microprocessors and ADCs are developed and also applied in nuclear instruments [2,3]. Sampling and recording the pulse waveform are becoming much easier and affordable. The advantages of using digital techniques in nuclear spectrum acquisition have widely been proved. These digitization techniques provide more possibilities than traditional analog techniques, such as enhancing signal-noise ratio, pile-up pulse correction and baseline correction. To get the nuclear spectrum from these digital waveform data, many methods to improve the nuclear spectrum have been developed[4, 5]. For example，one of them is to pick up the maximum value close to the peak position by simple comparing or polynomial fitting[6]. The main idea of these methods is to get the particle energy by the pulse height analysis and the accuracy of these methods is subject to the noise. But all these way need much more computation than traditional PHA.

In the traditional analog signal processing, pulse integration was used, such as the charge preamplifier is just the charge integration circuits [7], shaping circuit of semi-Gauss is also multiple integrals net, by these ways, high frequency noise can be eliminated or reduced effectively. In the present paper, a method of PAA after the amplifier based on digital pulse waveform is studied, by our knows, that is used for the first time.

In the method, to get the area of the pulse after amplifier digitized by high speed ADC, which is proportional to its amplitude, all the samples of each pulse are summed up. The main advantages of PAA are that it efficiently decreases the effect of the high frequency noise by averaging, thus can be used with a high accuracy at higher counting rate.

[*] Supported by National Natural Science Foundation of China (11375077,11027508)
[†] E-mail: pelab@lzu.edu.cn

Section 2 describes the principle and theory of PAA method, Section 3 describes the experiment realization and results of nuclear spectrum acquisition based on the digitizer and the traditional MCA. The conclusion is in the final Section.

**2 Method and simulation of PAA**

The output signal of the pulse amplifier is usually shaped to the quasi-Gauss waveform whose maximum amplitude represents the energy of the particle into the detector[8]. When the pulse shaper is CR-(RC)$^2$, the output waveform of the amplifier shown in Fig.1 can be expressed as

$$v_o(t) = At^2 e^{-t/\tau}, \quad t \geq 0 . \quad (1)$$

where A is the constant about amplitude and waveform, $\tau$ is time constant. Its peak value(PH) is proportional to the energy of the detected radiation and can be given as

$$PH = V_{om} = \frac{4A\tau^2}{e^2}, \quad t_m = 2\tau . \quad (2)$$

where $V_{om}$ is the amplitude of the pulse, and $t_m$ represents the time reaching the highest point. Taking into account the noise, the really PH is given by (3).

$$PH = V_{om} + v_n(2\tau) . \quad (3)$$

The PHA method is to obtain PH by ADC with pulse peak holder. Fig.2 is the reconstruction waveform sampling by the digitizer (U1066A-DC438, Agilent). When the details section of the sampling waveform are focused (the small figure in Fig.2), it is found that the relative fluctuation of data, which come from the high frequency noise, at the top of the pulse is closed to 1%. It severely restricts the resolution of PHA.

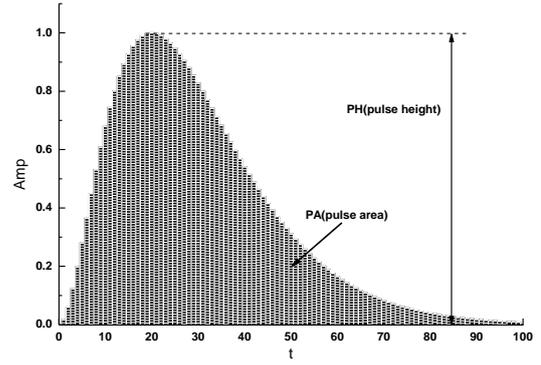

Fig.1 The semi-Gauss pulse waveform

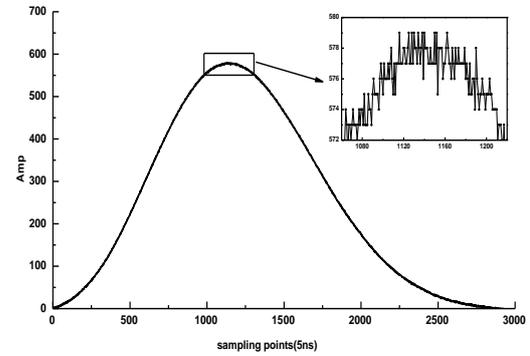

Fig.2 The pulse waveform sampled by the digitizer(U1066A-DC438, Agilent).

The pulse area(PA) is obtained by full integral to waveform as shown by (4).

$$PA = \int_0^\infty v_o(t)dt = 2A\tau^3 . \quad (4)$$

It is obvious that PA is proportional to the PH and the energy of the detected radiation. If the noise is taken into account, PA can be expressed as (5).

$$PA = \int_0^\infty v_o(t)dt + \int_0^\infty v_n(t)dt . \quad (5)$$

where $v_n(t)$ is noise. After digitizing by ADC, Formula (5) is converted to (6).

$$PA = \sum_{i=0}^m v_o(i) + \sum_{i=0}^m v_n(i) . \quad (6)$$

where m is the sampling points of waveform and PA is the adding of all the samples of the pulse. When m is much larger than one, we have

$$\sum_{i=0}^m v_n(i) \to 0 . \quad (7)$$

and then

$$PA \approx \sum_{i=0}^{m} v_o(i) . \tag{8}$$

By this simple sum way, the noise of fluctuation is dropped and the more precise results can be gotten.

Based on the above reasoning, the process of simulation is implemented by MatLab 7.0. The quasi-Gauss pulse without noise (pulse) and appending the noise (pulse+ noise) are shown in Fig.3 (a). To get more accurate results, 20000 pulses are generated for each process. The statistical distribution of PAA was gotten and fitted on different relative resolutions of PHA and different sampling number. The results of simulation are shown in Fig.3 (b). There are three relative resolutions of pulse height that were set $R_{PHA}$=10%, 1% and 0.1%. When sampling points are more than 1000 for each pulse, the relative resolution of PAA improves upon 1% from 10%, 0.1% from 1% and 0.01% from 0.1%. The noise is compressed and the resolution is improved for the signal with high frequency noise. The simulation shows that the resolution will be improved efficiently by the simple way.

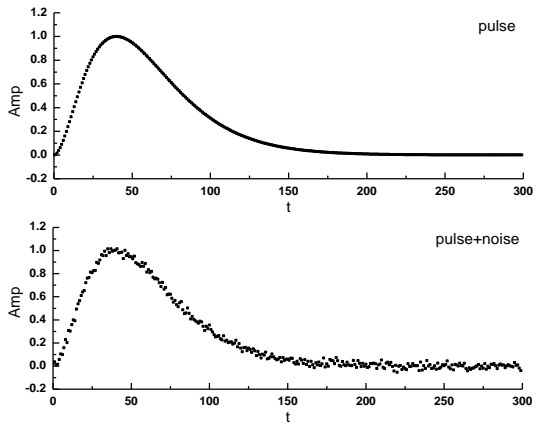

(a) Pulse waveform of simulation

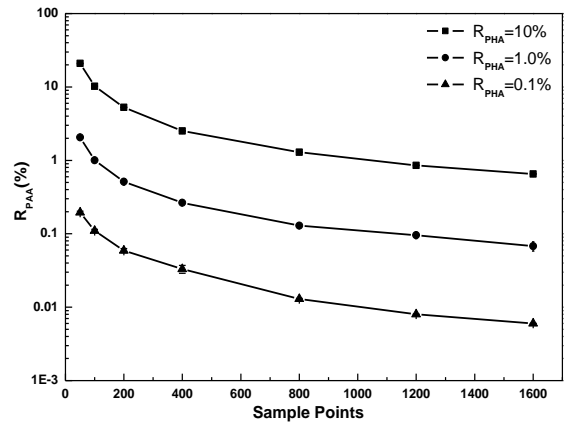

(b) Relative resolution of PAA

Fig. 3 Simulation results of PAA

## 3 Realization and experimental results

For verifying PAA, the digital acquisition system is built that is composed of a digitizer and a PC. With which the user can set the DPP(digital pulse process) parameters, choose the working mode and display the results by a VC program developed by our group(Control and Display software). As a contrast, traditional PHA is also testified with the same conditions.

The detector is high pure Germanium of ORTEC. To digitize the output pulses from the amplifier, a commercial digitizer (U1066A-DC438, Agilent)[9], with two high speed ADCs (12 bits and 200MS/s)and each channel with a 4M samples buffer memory, is employed. The data stream from each ADC is written in buffer memories and transported via NI PXI-8360 to PC. The MCA8000 (Amptek Inc.) based on PHA is employed too. The digital pulse processing and spectrum acquiring is carried out by the host computer.

In order to test the characteristic and feasibility of PAA, three isotope sources of $^{133}$Ba, $^{137}$Cs and $^{60}$Co were used. Energy range of γ ray is from 10keV to $10^3$keV. Fig.4 and Fig.5 are the obtained spectrum of these isotopes with MCA 8000 and digitizer separately.

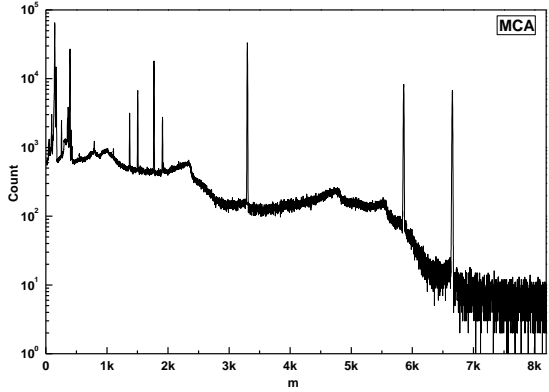

Fig. 4 The spectrum of MCA8000

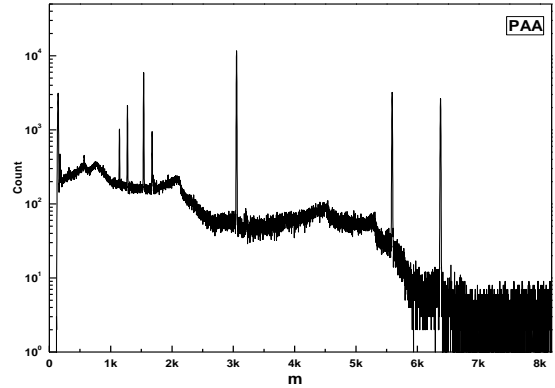

Fig. 5 The spectrum of digitizer

Since the ADC of the digitizer is only 12 bits, the spectrum of PHA has only 4096 channels. Nevertheless, as shown in Fig.6, the channel of spectrum based on PAA can be extended up to 32768 or higher by changing weighted coefficients. This method allows us to acquire high channel number spectrum using low bits ADC which is cheaper and faster than high bits one.

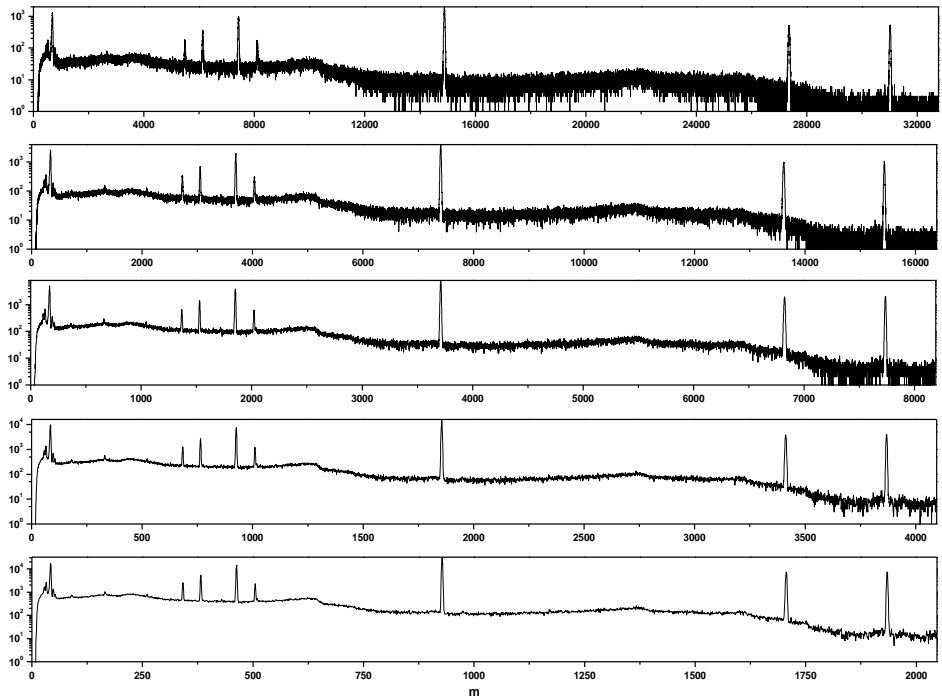

Fig. 6 The spectrum of PAA with different maximum channel

With the shaping time constant changing from 0.5us to 2us and other conditions unchanged, we acquired the spectra by both of digitizer and MCA8000 and calculated FWHMs of 5 chosen typical peaks. The results are shown in table 1. When time constant is 0.5us and 1.0us, the resolution of PAA is better than that of PHA. As time constant is 2.0us, the resolution of PAA and PHA doesn't show significant differences. As we all know, the time constant is an compromise between high count rate and high resolution. Through the simple PAA, the better resolution will be achieved with the small time constant and this method is applicable to high count rate.

**Table 1** Energy resolution of a digitizer spectrometer(DIG-PAA) and a classical spectrometer (MCA-PHA) in different time constant(T.C.) of shaping.

| ENERGY (keV) | 0.5μs FWHM$_E$(keV) | | 1.0μs FWHM$_E$(keV) | | 2.0μs FWHM$_E$(keV) | |
|---|---|---|---|---|---|---|
| | MCA-PHA | DIG-PAA | MCA-PHA | DIG-PAA | MCA-PHA | DIG-PAA |
| 80.999 | 1.944 | 1.559 | 1.351 | 1.132 | 0.979 | 0.980 |
| 356.014 | 3.197 | 1.900 | 1.805 | 1.603 | 1.311 | 1.550 |
| 661.661 | 5.014 | 2.602 | 2.247 | 1.784 | 1.600 | 1.651 |
| 1173.24 | 8.340 | 4.927 | 3.162 | 2.149 | 2.067 | 2.035 |
| 1332.51 | 9.168 | 5.861 | 3.498 | 2.216 | 2.249 | 2.160 |

For our further study the resolutions obtained with PAA and PHA, we chose $^{133}$Ba as γ source, since it has four characteristic γ rays with close energy as shown in Table 2. It can be seen that, the FWHMs and relative resolutions are much better for PAA than for PHA at the smaller shaping time constant. With increasing the shaping time constant, the resolution of PHA becomes close to that of PAA does. The mean relative resolutions were also calculated and shown in Table 2. It is very clear that when T.C. is 0.5μs, the resolution of PAA is about 66% better than that of PHA, but when T.C. is 2.0μs, both resolutions don't have significant difference.

**Table 2** Relative resolutions of PHA and PAA for $^{133}$Ba in different time constants of shaping

| T.C. ENERGY (keV) | 0.5μs MCA-PHA $\eta_1$(%) | DIG-PAA $\eta_2$(%) | 1.0μs MCA-PHA $\eta_1$(%) | DIG-PAA $\eta_2$(%) | 2.0μs MCA-PHA $\eta_1$(%) | DIG-PAA $\eta_2$(%) |
|---|---|---|---|---|---|---|
| 276.404 | 0.977 | 0.600 | 0.565 | 0.471 | 0.412 | 0.435 |
| 302.858 | 0.923 | 0.559 | 0.531 | 0.423 | 0.391 | 0.389 |
| 356.014 | 0.867 | 0.529 | 0.466 | 0.372 | 0.353 | 0.334 |
| 383.859 | 0.831 | 0.477 | 0.424 | 0.351 | 0.318 | 0.301 |
| $\bar{\eta}$ | 0.900% | 0.541% | 0.497% | 0.404% | 0.369% | 0.365% |

**4 Conclusions**

A new simple method of acquisition nuclear spectrum based on PAA proposed in this paper was proved, tested and evaluated with the HPGe detector. It turns out that the method can effectively reduce the effect of the noise to improve the energy resolution. The obtained results show that when the shaping time constant is 0.5μs, the energy resolution of PAA can be 66% better than that of PHA. For high count rate, the proposed method is a promising way to improve the energy resolution.

Acknowledgement: The authors would like to acknowledge the support from the National Natural Science Foundation of China with Grant No.11375077,